\documentclass[12pt]{article}
\usepackage{graphicx}
\usepackage{lscape}
\usepackage{citesort}
\usepackage{amssymb}
\usepackage{appendix}
\usepackage{multirow}

\begin{document}
\def\qq{\langle \bar q q \rangle}
\def\uu{\langle \bar u u \rangle}
\def\dd{\langle \bar d d \rangle}
\def\sp{\langle \bar s s \rangle}
\def\GG{\langle g_s^2 G^2 \rangle}
\def\Tr{\mbox{Tr}}
\def\figt#1#2#3{
        \begin{figure}
        $\left. \right.$
        \vspace*{-2cm}
        \begin{center}
        \includegraphics[width=10cm]{#1}
        \end{center}
        \vspace*{-0.2cm}
        \caption{#3}
        \label{#2}
        \end{figure}
	}
	
\def\figb#1#2#3{
        \begin{figure}
        $\left. \right.$
        \vspace*{-1cm}
        \begin{center}
        \includegraphics[width=10cm]{#1}
        \end{center}
        \vspace*{-0.2cm}
        \caption{#3}
        \label{#2}
        \end{figure}
                }

\def\ds{\displaystyle}
\def\beq{\begin{equation}}
\def\eeq{\end{equation}}
\def\bea{\begin{eqnarray}}
\def\eea{\end{eqnarray}}
\def\beeq{\begin{eqnarray}}
\def\eeeq{\end{eqnarray}}
\def\ve{\vert}
\def\vel{\left|}
\def\ver{\right|}
\def\nnb{\nonumber}
\def\ga{\left(}
\def\dr{\right)}
\def\aga{\left\{}
\def\adr{\right\}}
\def\lla{\left<}
\def\rra{\right>}
\def\rar{\rightarrow}
\def\lrar{\leftrightarrow}  
\def\nnb{\nonumber}
\def\la{\langle}
\def\ra{\rangle}
\def\ba{\begin{array}}
\def\ea{\end{array}}
\def\tr{\mbox{Tr}}
\def\ssp{{\Sigma^{*+}}}
\def\sso{{\Sigma^{*0}}}
\def\ssm{{\Sigma^{*-}}}
\def\xis0{{\Xi^{*0}}}
\def\xism{{\Xi^{*-}}}
\def\qs{\la \bar s s \ra}
\def\qu{\la \bar u u \ra}
\def\qd{\la \bar d d \ra}
\def\qq{\la \bar q q \ra}
\def\gGgG{\la g^2 G^2 \ra}
\def\q{\gamma_5 \not\!q}
\def\x{\gamma_5 \not\!x}
\def\g5{\gamma_5}
\def\sb{S_Q^{cf}}
\def\sd{S_d^{be}}
\def\su{S_u^{ad}}
\def\sbp{{S}_Q^{'cf}}
\def\sdp{{S}_d^{'be}}
\def\sup{{S}_u^{'ad}}
\def\ssp{{S}_s^{'??}}

\def\sig{\sigma_{\mu \nu} \gamma_5 p^\mu q^\nu}
\def\fo{f_0(\frac{s_0}{M^2})}
\def\ffi{f_1(\frac{s_0}{M^2})}
\def\fii{f_2(\frac{s_0}{M^2})}
\def\O{{\cal O}}
\def\sl{{\Sigma^0 \Lambda}}
\def\es{\!\!\! &=& \!\!\!}
\def\ap{\!\!\! &\approx& \!\!\!}
\def\ar{&+& \!\!\!}
\def\ek{&-& \!\!\!}
\def\kek{\!\!\!&-& \!\!\!}
\def\cp{&\times& \!\!\!}
\def\se{\!\!\! &\simeq& \!\!\!}
\def\eqv{&\equiv& \!\!\!}
\def\kpm{&\pm& \!\!\!}
\def\kmp{&\mp& \!\!\!}
\def\mcdot{\!\cdot\!}
\def\erar{&\rightarrow&}


\def\simlt{\stackrel{<}{{}_\sim}}
\def\simgt{\stackrel{>}{{}_\sim}}


\title{
         {\Large
                 {\bf
Analysis of $\gamma^\ast \Lambda \to \Sigma^0$ transition in QCD 
                 }
         }
      }

\author{\vspace{1cm}\\
{\small T. M. Aliev \thanks {e-mail:
taliev@metu.edu.tr}~\footnote{permanent address:Institute of
Physics,Baku,Azerbaijan}\,\,, K. Azizi \thanks {e-mail:
kazizi@dogus.edu.tr}\,\,, M. Savc{\i} \thanks
{e-mail: savci@metu.edu.tr}} \\
{\small Physics Department, Middle East Technical University,
06531 Ankara, Turkey }\\
{\small$^\ddag$ Physics Department,  Faculty of Arts and Sciences,
Do\u gu\c s University,} \\
{\small Ac{\i}badem-Kad{\i}k\"oy,  34722 Istanbul, Turkey}}

\date{}

\begin{titlepage}
\maketitle
\thispagestyle{empty}

\begin{abstract}

The $\gamma^\ast \Lambda \to \Sigma^0$ transition form factors are
investigated within the light--cone QCD sum rules method. Using the most
general form of the interpolating current of $\Sigma^0$ baryon and the
distribution amplitudes of $\Lambda$ baryon we calculate the $Q^2$
dependence of the electromagnetic form factors. Our result are compared
with the predictions of the covariant spectator quark model.

\end{abstract}

~~~PACS numbers: 11.55.Hx, 12.38.--t, 13.40.Gp

\end{titlepage}

\section{Introduction}

The investigation of electromagnetic form factors of hadrons plays a key role in
understanding their internal structure. The form factors measured in
experiments describe the spatial distribution of charge and magnetization of
hadrons \cite{Rslff01}, and indicate the deviation of hadron structure
from the point-like particle. At present, the studies are
mainly focused on the nucleon form factors. Recent experimental and
theoretical progress on this subject can be found in \cite{Rslff01,Rslff02}
and references therein.

The study of electromagnetic form factors of the ground state spin--1/2
baryons receives special interest. However, except the proton and neutron,
the electromagnetic form factors of other members the octet baryons have not
yet been measured. The main difficulty can be attributed to the unstable
nature of the baryons containing strange quarks. From theoretical point of
view, the main problem is related to the fact that the formation of
hadrons belong to the nonperturbative region of QCD where perturbative
approach does not work. For this reason some nonperturbative approaches are
needed in order to calculate these form factors, and the QCD sum rules
method is recognized to be the most predictive one among all other
nonperturbative approaches. Another advantage of the QCD sum rules method is
that it is based on the fundamental QCD Lagrangian.

The nucleon electromagnetic form factors are calculated in the framework of the
light-cone version of  QCD sum rules method for the Ioffe and general currents
in \cite{Rslff03} and \cite{Rslff04}. The electromagnetic form factors of
$\Lambda$, $\Sigma$ and $\Xi$ baryons are studied for the
Chernyak--Zhitnisky and Ioffe currents in \cite{Rslff05}. The
electromagnetic form factors of octet baryons for the most general form of
the interpolating currents, are studied within the light--cone QCD sum rules
method in \cite{Rslff06}. It should be noted here that, the electromagnetic form
factors of nucleons and other members of octet baryons have already been
studied in numerous works within the framework of lattice calculations
(see \cite{Rslff07} and references therein), and relativistic constituent
quark model in \cite{Rslff08}.

In the present work we study the electromagnetic transition form factors of the $\gamma^\ast \Lambda \to \Sigma^0$ in the 
framework of the light--cone QCD sum rules method using the most general
form of the interpolating current for the $\Sigma^0$ baryon. This decay is
studied in framework of the nonrelativistic quark model and general QCD parametrization method \cite{Morpurgo}, the covariant spectator quark model \cite{Rslff09},
chiral perturbative theory \cite{Rslff10,Rslff11}, chiral quark model
\cite{Rslff12} and Skyrme model \cite{Rslff13}. The $\gamma^\ast \Lambda \to
\Sigma^0$ transition is interesting in several respects: it is
unique between two different baryons that belong to the same octet family
even in exact isospin symmetry case. The second interesting peculiarity of
this transition is that having different initial and final baryons is
contrary to the case observed in elastic scattering of the octet baryons. 
For these reasons, the electric charge form factor $G_E(Q^2)$ at $Q^2=0$
should vanish. Hence, the value of $G_E(Q^2)$ is expected to be small in its
dependence on $Q^2$. Therefore, investigation of the $Q^2$ dependence of 
the form factors receives special interest. It should be noted that the
magnetic moment for the $\gamma^\ast \Lambda \to \Sigma^0$ transition is
investigated within the light--cone QCD sum rules method in \cite{Rslff14}. 
The modern status of QCD and particularly the  QCD sum rules for baryons is  presented in great detail in \cite{Ioffebook}.

The structure of this paper is organized as follows. In Section 2 we derive
sum rules for the form factors of the $\gamma^\ast \Lambda \to 
\Sigma^0$ transition. In Section 3 we present our numerical results
and conclusions.   

\section{Sum rules for $\gamma^\ast \Lambda \to \Sigma^0$ transition form
factors}

The transition form factors for $\gamma^\ast \Lambda \to \Sigma^0$
is determined by the matrix element of the electromagnetic current between
the $\Lambda$ and $\Sigma^0$ baryons. Using the conservation of
electromagnetic current, this matrix element can be determined in the
following way:

\bea
\label{eslff01}
\lla \Sigma^0(p^\prime) \vel j_\mu^{el} \ver \Lambda(p) \rra =
\bar{u}_{\Sigma^0}(p^\prime) \Bigg\{ F_1(Q^2) \Bigg( \gamma_\mu - {\rlap/{q} q_\mu
\over q^2} \Bigg) - {i\over m_\Lambda + m_{\Sigma^0}} \sigma_{\mu\nu} q^\nu
F_2(Q^2) \Bigg\} u_\Lambda(p)~,
\eea
where $q=p-p^\prime$, $Q^2=-q^2$ and $\sigma_{\mu\nu} = {i\over 2}
\left[\gamma_\mu,\gamma_\nu\right]$. Here, $F_1(Q^2)$ and $F_2(Q^2)$ are the
Dirac and Pauli type form factors, respectively.

Experimentally, more convenient set of the electromagnetic form factors are
the Saches form factors defined as,

\bea
\label{eslff02} 
G_E(Q^2) \es F_1(Q^2) - {Q^2\over (m_\Lambda + m_{\Sigma^0})^2} F_2(Q^2)~, \nnb \\
G_M(Q^2) \es F_1(Q^2) + F_2(Q^2)~.
\eea

In order to calculate the form factors $F_1(Q^2)$ and $F_2(Q^2)$ for the
$\gamma^\ast \Lambda \to \Sigma^0$ transition we consider the following
correlation function:
\bea
\label{eslff03}
\Pi_\mu(p,q) = i \int d^4x e^{iqx} \lla 0 \vel \mbox{T} \Big\{ \eta^{\Sigma_0}(0)
j_\mu^{el} (x) \Big\} \ver\Lambda(p) \rra~,
\eea
where T means the time ordering, $\left. \left. \ver \Lambda(p) \rra$ is the
$\Lambda$ baryon state with four--momentum $p$, $\eta^{\Sigma_0}$ is the
interpolating current for the $\Sigma_0$ baryon, i.e., 
\bea
\label{eslff04}
\eta^{\Sigma_0} \es \sqrt{2} \varepsilon^{abc} \Big\{ ( u^{aT} C s^b)
\gamma_5 d^c + ( d^{aT} C s^b) \gamma_5 u^c \nnb \\
\ar \beta ( u^{aT} C \gamma_5 s^b) d^c + \beta ( d^{aT} C \gamma_5 s^b) u^c
\Big\}~.
\eea
Here $C$ is the charge conjugation operator,  $\beta$ is an arbitrary
parameter and $j_\mu^{el}$ is the electromagnetic current defined as
\bea
\label{eslff05}
j_\mu^{el} (x) = e_u \bar{u}(x)\gamma_\mu u(x) + e_d \bar{d}(x)\gamma_\mu d(x) +
 e_s \bar{s}(x)\gamma_\mu s(x)~.
\eea

The correlation function can be calculated in terms of hadrons
(phenomenological part) and in terms of quark and gluon degrees of freedom.
Equating these two representations of the correlation function
(\ref{eslff01}) we get the sum rules for the form factors of $\gamma^\ast
\Lambda \to \Sigma^0$ transition.

Saturating (\ref{eslff01}) with the hadronic states with the quantum numbers
of $\Sigma^0$ baryon and separating the ground state, for the phenomenological
part we get
\bea
\label{eslff06}
\Pi_\mu(p,q) = {\ds \lla 0 \vel \eta^{\Sigma_0} \ver \Sigma_0(p^{'}) \rra
\lla \Sigma_0(p^{'}) \vel j_\mu^{el} \ver \Lambda (p) \rra \over
 m_{\Sigma^0}^2 - p^{\prime 2} } + \cdots~,
\eea
where $\cdots$ denotes  contribution of the  higher states and continuum.

The matrix element $\lla 0 \vel \eta^{\Sigma_0} \ver \Sigma_0 \rra$ is
determined as
\bea
\label{nolabel01}
\lla 0 \vel \eta^{\Sigma_0} \ver \Sigma_0 \rra = \lambda_{\Sigma_0}
u(p^\prime)~, \nnb
\eea
where $\lambda_{\Sigma_0}$ is the residue of $\Sigma_0$ baryon. Moreover,
the matrix element $\lla \Sigma_0 \vel j_\mu^{el} \ver \Lambda (p) \rra$ is
determined as is given in Eq. (\ref{eslff01}).
Using these definitions, for the phenomenological part we get
\bea
\label{eslff07}
\Pi_\mu^{ph} = {\lambda_{\Sigma_0} (\rlap/{p}^\prime + m_{\Sigma^0}) \over
 m_{\Sigma^0}^2 - p^{\prime 2} } \Bigg\{ F_1(Q^2) \Bigg( \gamma_\mu -
{\rlap/{q} q_\mu
\over q^2} \Bigg) - {i\over m_\Lambda + m_{\Sigma^0}} \sigma_{\mu\nu} q^\nu
F_2(Q^2) \Bigg\} u_\Lambda(p)~.
\eea
We see from Eq. (\ref{eslff07}) that there appears numerous structures in
determining the transition form factors $F_1(Q^2)$ and $F_2(Q^2)$. For this
aim we choose the structures $p_\mu$ and $p_\mu \rlap/{q}$, as a result of which, for the coefficients of the selected structures, we get
\bea
\label{eslff08}
\Pi^{(1)} \es {2 \lambda_{\Sigma_0} F_1(Q^2) \over m_{\Sigma^0}^2 -
p^{\prime 2}}~, \nnb \\
\Pi^{(2)} \es  {2 \over m_{\Sigma_0} + m_\Lambda}{\lambda_{\Sigma_0} F_2(Q^2) \over m_{\Sigma^0}^2 -         
p^{\prime 2}}~.
\eea 

As has already been noted, these form
factors are described in terms of $\Lambda$ baryon distribution
amplitudes (DAs). The $\Lambda$ baryon
matrix element of three--quark operator $ \varepsilon^{abc} \lla \left.
u_\alpha^a(a_1 x) d_\beta^b(a_2 x) s_\gamma^c (a_3 x) \ver \Lambda(p) \rra$ 
is given in terms of $\Lambda$ baryon DAs.The definition of this matrix
element in terms of DAs, and expressions of these DAs can be found
\cite{Rslff05}.

In constructing sum rules for the transition form factors $F_1(Q^2)$
and $F_2(Q^2)$ we need the expression for the correlation function from the
QCD side. This correlation function in QCD can be calculated for large
negative $p^{'2}$ and $q^2=-Q^2$ in terms of $\Lambda$ baryon distribution
amplitudes using the operator product expansion. Matching then the
coefficients of the structures $p_\mu$ and $p_\mu \rlap/{p}$ in the
expressions of the correlation function in the phenomenological and QCD sides,
we get the sum rules for the transition form factors $F_1(Q^2)$ and
$F_2(Q^2)$ of the $\gamma^\ast \Lambda \to \Sigma^0$ transition.
  
In order to enhance the
ground state contribution and suppress the higher state contributions, it is
necessary to perform Borel transformation on theoretical and
phenomenological parts of the correlation function. After the Borel
transformation we get the final expressions for the transition form factors
$F_1(Q^2)$ and $F_2(Q^2)$ as

\bea
\label{eslff09}
F_1(Q^2) \es {\sqrt{2} \over 4} {1\over 2 \lambda_{\Sigma^0}}
e^{m_{\Sigma^0}^2/M^2} \Bigg\{ \int_{x_0}^1 dx \left(
- {\rho_2(x)\over x} + {\rho_4(x)\over M^2 x^2} - {\rho_6(x)\over 2 M^4 x^3 } \right)
e^{-\left( {Q^2 \bar{x} \over M^2 x} + {m_{\Lambda}^2 \bar{x} \over
M^2}\right)} \nnb \\
\ar \Bigg[ {\rho_4 (x_0) \over Q^2 + m_{\Lambda}^2 x_0^2} - {1\over 2 x_0}
{\rho_6(x_0) \over (Q^2 + m_{\Lambda}^2 x_0^2) M^2} \nnb \\
\ar {1\over 2} {x_0^2 \over (Q^2 + m_{\Lambda}^2 x_0^2)} \left(
{d\over dx_0} {\rho_6(x_0) \over x_0 (Q^2 + m_{\Lambda}^2 x_0^2) M^2}
\right) \Bigg]e^{-s_0/ M^2}
\Bigg\}~, \\ \nnb \\ \nnb \\
\label{eslff10}
F_2(Q^2) \es {\sqrt{2} \over 4} {m_{\Sigma^0} + m_\Lambda \over 2 \lambda_{\Sigma^0}}
e^{m_{\Sigma^0}^2/M^2}
\Bigg\{ \int_{x_0}^1 dx \left(
- {\rho^{'}_2(x)\over x} + {\rho^{'}_4(x)\over M^2 x^2} - {\rho^{'}_6(x)\over 2 M^4 x^3} \right)
e^{-\left( {Q^2 \bar{x} \over M^2 x} + {m_{\Lambda}^2 \bar{x} \over
M^2}\right)} \nnb \\
\ar \Bigg[ {\rho^{'}_4 (x_0) \over Q^2 + m_{\Lambda}^2 x_0^2} - {1\over 2 x_0}
{\rho^{'}_6(x_0) \over (Q^2 + m_{\Lambda}^2 x_0^2) M^2} \nnb \\
\ar {1\over 2} {x_0^2 \over (Q^2 + m_{\Lambda}^2 x_0^2)} \left(
{d\over dx_0} {\rho^{'}_6(x_0) \over x_0 (Q^2 + m_{\Lambda}^2 x_0^2) M^2}
\right) \Bigg]e^{-s_0/ M^2}
\Bigg\}~,
\eea
where,
\bea
\label{eslff11}
\rho_6(x) \es
4 e_u m_\Lambda^3 (1+\beta) x (m_\Lambda^2 x^2 + Q^2)
\;\check{\!\check{B}}_6 (x)
+ 4 e_d m_\Lambda^3 (1+\beta) x (m_\Lambda^2 x^2 + Q^2)
\;\widetilde{\!\widetilde{B}}_6 (x) \nnb \\
\ar 8 e_s m_\Lambda^2 \Big\{ m_\Lambda^2 m_s (1-\beta) x^2\; \widehat{\!\widehat{C}}_6
+ (1+\beta) \Big[ m_\Lambda x (m_\Lambda^2 x^2 + Q^2)
\; \widehat{\!\widehat{B}}_6 \nnb \\
\ek m_s ( Q^2 \; \widehat{\!\widehat{B}}_6 + 2 m_\Lambda^2 x^2
\;\widehat{\!\widehat{B}}_8) \Big]\Big\} (x)~, \nnb \\ \nnb \\
\rho_4 (x)\es
e_u m_\Lambda \Big\{
- 2 m_\Lambda^2 x \Big[ 2 (1-\beta) \; \check{\!\check{C}}_6
 - (1+\beta) (2 \; \check{\!\check{B}}_6 - 5 \;\check{\!\check{B}}_8) \Big] (x) \nnb \\
\ar \Big[ 2 (1-\beta) \Big( m_\Lambda^2 x^2
(\check{D}_5
- \check{C}_4 + 2 \check{C}_5 )
- Q^2 (\check{D}_2 - \check{C}_2) \Big) \nnb \\
\ar (1+\beta) \Big( Q^2 (3 \check{B}_2 + 7                   
\check{B}_4)  + m_\Lambda^2 x^2 ( 2 \check{H}_1 
- 2 \check{E}_1 - \check{B}_2 +                
\check{B}_4 
- 10 \check{B}_5 - 20 \check{B}_7 ) \Big) \Big] (x) \nnb \\
\ek 2 m_\Lambda^2 x \int_0^{\bar{x}}dx_3\,\Big[ 2 (1-\beta) {V}_1^M + 5 (1+\beta)
T_1^M\Big] (x,1-x-x_3,x_3) \Big\}\nnb \\
\ar e_d m_\Lambda \Big\{
- 2 m_\Lambda^2 x \Big[ 2 (1-\beta) \; \widetilde{\!\widetilde{C}}_6
 - (1+\beta) (2 \; \widetilde{\!\widetilde{B}}_6 - 
5 \; \widetilde{\!\widetilde{B}}_8) \Big] (x) \nnb \\
\ar \Big[ (1-\beta) \Big( - 2 m_\Lambda^2 x^2 (\widetilde{D}_5
+ \widetilde{C}_4 - 2 \widetilde{C}_5 )  
+ Q^2 (\widetilde{D}_2 + \widetilde{C}_2) \Big) \nnb \\
\ar (1+\beta) \Big( Q^2 (3 \widetilde{B}_2 + 7
\widetilde{B}_4)  - m_\Lambda^2 x^2 ( 2 \widetilde{H}_1
- 2 \widetilde{E}_1 + \widetilde{B}_2 -
\widetilde{B}_4
+  10 \widetilde{B}_5 + 20 \widetilde{B}_7 ) \Big) \Big] (x) \nnb \\
\ek  2 m_\Lambda^2 x \int_0^{\bar{x}} \,dx_1 \Big[ 2 (1-\beta) {V}_1^M + 5 (1+\beta)
T_1^M\Big] (x_1,x,1-x_1-x) \Big\}\nnb \\
\ar 2 e_s m_\Lambda \Big\{
2 m_\Lambda (1+\beta) \Big[ m_\Lambda x
(2  \; \widehat{\!\widehat{B}}_6 - \widehat{\!\widehat{B}}_8 ) - m_s \;
\widehat{\!\widehat{B}}_6 \Big] (x) \nnb \\
\ar \Big[ (1-\beta) \Big(
2 (m_\Lambda^2 x^2 \widehat{C}_5 + Q^2 \widehat{C}_2) -
m_\Lambda m_s x ( 2 \widehat{C}_2 - \widehat{C}_4 -
\widehat{C}_5 ) \Big) \nnb \\
\ek (1+\beta) \Big( Q^2 ( \widehat{B}_2 - 3 \widehat{B}_4) +
m_\Lambda^2 x^2 ( \widehat{B}_2 - \widehat{B}_4 + 2 \widehat{B}_5 +
4 \widehat{B}_7)
- 4 m_\Lambda m_s x ( \widehat{B}_4 -
\widehat{B}_5)\Big) \Big] (x) \nnb \\
\ek 2m_\Lambda^2 (1+\beta) x \int_0^{\bar{x}}dx_1\, T_1^M (x_1,1-x_1-x,x) \Big\}~, \nnb \\ \nnb \\
\rho_2(x) \es
 - 2 e_u m_\Lambda \Big\{
\Big[ (1-\beta) ( \check{D}_2 + \check{C}_2) -
(1+\beta) ( \check{B}_2 - \check{B}_4) \Big] (x) \nnb \\
\ar x \int_0^{\bar{x}}dx_3\,\Big[ (1-\beta) (A_3 + 2 {V}_1 - 3 {V}_3) -
(1+\beta)(P_1 + S_1 - 5 T_1 + 10 T_3) \Big] (x,1-x-x_3,x_3) \Big\} \nnb \\
\ar 2 e_d m_\Lambda \Big\{
\Big[ (1-\beta) ( \widetilde{D}_2 - \widetilde{C}_2) +
(1+\beta) ( \widetilde{B}_2 - \widetilde{B}_4) \Big] (x) \nnb \\
\ar x \int_0^{\bar{x}} \,dx_1 \Big[ (1-\beta) (A_3 - 2 {V}_1 + 3 {V}_3) -
(1+\beta)(P_1 + S_1 + 5 T_1 - 10 T_3) \Big] (x_1,x,1-x_1-x) \Big\} \nnb \\
\ar 4 e_s \Big\{
m_\Lambda \Big[ (1-\beta) \widehat{C}_2 -  
(1+\beta)  ( \widehat{B}_2 - \widehat{B}_4) \Big] (x) \nnb \\
\ar \int_0^{\bar{x}}dx_1\,\Big\{ (1-\beta) ( m_\Lambda x {V}_3 + m_s {V}_1 )
+ (1+\beta) \Big[ 2 m_\Lambda x T_3 - (m_\Lambda x + 2 m_s) T_1 \Big]
\Big\} (x_1,1-x_1-x,x)\Big\}~, \nnb \\ \nnb \\
\rho_6^{'} (x) \es
 - 4 e_u m_\Lambda^2 (1+\beta) (m_\Lambda^2 x^2 + Q^2)  
\;\check{\!\check{B}}_6 (x)
- 4 e_d m_\Lambda^2 (1+\beta) (m_\Lambda^2 x^2 + Q^2)  
\;\widetilde{\!\widetilde{B}}_6 (x) \nnb \\
\ek 8 e_s m_\Lambda^2  \Big\{ m_\Lambda m_s (1-\beta) x \; \widehat{\!\widehat{C}}_6
+ (1+\beta) \Big[ (m_\Lambda^2 x^2 + Q^2) \; \widehat{\!\widehat{B}}_6 +
m_\Lambda m_s x (\; \widehat{\!\widehat{B}}_6 - 2 \; \widehat{\!\widehat{B}}_8 )
\Big]\Big\} (x)~, \nnb \\ \nnb \\
\rho_4^{'} (x) \es
- e_u m_\Lambda^2 \Big\{
(1+\beta) \;\check{\!\check{B}}_6 (x) \nnb \\
\ar 2 x \Big[ (1-\beta) (\check{D}_2 + \check{D}_5 - \check{C}_2 -
\check{C}_4 + 2 \check{C}_5 ) +
(1+\beta) (\check{H}_1 - \check{E}_1 - 2 \check{B}_2 - 3 \check{B}_4 - 5
\check{B}_5 - 10 \check{B}_7 ) \Big] (x) \nnb \\
\ar 2 (1-\beta) \int_0^{\bar{x}}dx_3\, (A_1^M - {V}_1^M) (x,1-x-x_3,x_3)\Big\} \nnb \\
\ar e_d m_\Lambda^2 \Big\{ 
- (1+\beta) \;\widetilde{\!\widetilde{B}}_6 (x) \nnb \\
\ar 2 x \Big[ (1-\beta) (\widetilde{D}_2 + \widetilde{D}_5 + \widetilde{C}_2 + 
\widetilde{C}_4 - 2 \widetilde{C}_5 ) +
(1+\beta) (\widetilde{H}_1 - \widetilde{E}_1 + 2 \widetilde{B}_2 + 3 \widetilde{B}_4 
+ 5 \widetilde{B}_5 + 10 \widetilde{B}_7 ) \Big] (x) \nnb \\
\ar 2  (1-\beta) \int_0^{\bar{x}}dx_1 \, (A_1^M + {V}_1^M) (x_1,x,1-x_1-x)\Big\} \nnb \\
\ek 2 e_s  m_\Lambda \Big\{ 
5 m_\Lambda (1+\beta) \;\widehat{\!\widehat{B}}_6 (x) \nnb \\
\ar 2 \Big[ (1-\beta) \Big( m_\Lambda x \widehat{C}_5 - 
( m_\Lambda x + m_s) \widehat{C}_2 \Big) -
(1+\beta) \Big( m_\Lambda x (\widehat{B}_4 + \widehat{B}_5 + 2 \widehat{B}_7) -
m_s ( \widehat{B}_2 + \widehat{B}_4) \Big) \Big] (x) \nnb \\
\ar 2 m_\Lambda (1-\beta) \int_0^{\bar{x}}dx_1\, {V}_1^M (x_1,1-x_1-x,x)\Big\}~, \nnb \\ \nnb \\
\rho_2^{'} (x) \es - 2 e_u (1-\beta) \int_0^{\bar{x}}dx_3\, (A_1 - {V}_1) (x,1-x-x_3,x_3) \nnb \\
\ar 2 e_d (1-\beta) \int_0^{\bar{x}}dx_1 \,(A_1 + {V}_1) (x_1,x,1-x_1-x) \nnb \\
\ek 4 e_s (1-\beta) \int_0^{\bar{x}}dx_1\,{V}_1 (x,1-x-x_3,x_3)~, 
\eea
where $M^2$ is the Borel parameter and $x_0$ is given as
\bea
\label{nolabel}
x_0 = {\sqrt{(Q^2+ s_0 - m_\Lambda^2)^2 + 4 m_\Lambda^2 Q^2}\over 2
m_\Lambda^2}~. \nnb
\eea
Here $s_0$ is the continuum threshold.
In the expressions of $\rho_i^{(')}(x)$,  the functions ${\cal F}(x_i)$ are defined as

\bea
\label{eslff12}
\check{\cal F}(x_1) \es \int_1^{x_1}\!\!dx_1^{'}\int_0^{1- x^{'}_{1}}\!\!dx_3\,
{\cal F}(x_1^{'},1-x_1^{'}-x_3,x_3)~, \nnb \\
\check{\!\!\!\;\check{\cal F}}(x_1) \es 
\int_1^{x_1}\!\!dx_1^{'}\int_1^{x^{'}_{1}}\!\!dx_1^{''}
\int_0^{1- x^{''}_{1}}\!\!dx_3\,
{\cal F}(x_1^{''},1-x_1^{''}-x_3,x_3)~, \nnb \\
\widetilde{\cal F}(x_2) \es \int_1^{x_2}\!\!dx_2^{'}\int_0^{1- x^{'}_{2}}\!\!dx_1\,
{\cal F}(x_1,x_2^{'},1-x_1-x_2^{'})~, \nnb \\
\widetilde{\!\widetilde{\cal F}}(x_2) \es 
\int_1^{x_2}\!\!dx_2^{'}\int_1^{x^{'}_{2}}\!\!dx_2^{''}
\int_0^{1- x^{''}_{2}}\!\!dx_1\,
{\cal F}(x_1,x_2^{''},1-x_1-x_2^{''})~, \nnb \\
\widehat{\cal F}(x_3) \es \int_1^{x_3}\!\!dx_3^{'}\int_0^{1- x^{'}_{3}}\!\!dx_1\,
{\cal F}(x_1,1-x_1-x_3^{'},x_3^{'})~, \nnb \\
\widehat{\!\widehat{\cal F}}(x_3) \es 
\int_1^{x_3}\!\!dx_3^{'}\int_1^{x^{'}_{3}}\!\!dx_3^{''}
\int_0^{1- x^{''}_{3}}\!\!dx_1\,
{\cal F}(x_1,1-x_1-x_3^{''},x_3^{''})~.
\eea

We also use the following
shorthand notations for the combinations of the distribution amplitudes:

\bea
\label{eslff13}
B_2 \es T_1+T_2-2 T_3~, \nnb \\
B_4 \es T_1-T_2-2 T_7~, \nnb \\
B_5 \es - T_1+T_5+2 T_8~, \nnb \\
B_6 \es 2 T_1-2 T_3-2 T_4+2 T_5+2 T_7+2 T_8~, \nnb \\
B_7 \es T_7-T_8~, \nnb \\
B_8 \es  -T_1+T_2+T_5-T_6+2 T_7+2T_8~, \nnb \\
C_2 \es V_1-V_2-V_3~, \nnb \\
C_4 \es -2V_1+V_3+V_4+2V_5~, \nnb \\
C_5 \es V_4-V_3~, \nnb \\
C_6 \es -V_1+V_2+V_3+V_4+V_5-V_6~, \nnb \\
D_2 \es -A_1+A_2-A_3~, \nnb \\
D_4 \es -2A_1-A_3-A_4+2A_5~, \nnb \\
D_5 \es A_3-A_4~, \nnb \\
D_6 \es A_1-A_2+A_3+A_4-A_5+A_6~, \nnb \\
E_1 \es S_1-S_2~, \nnb \\
H_1 \es P_2-P_1~.
\eea

It follows from Eqs. (\ref{eslff09}) and (\ref{eslff10}) that in order to calculate the form factors $F_1(Q^2)$
and $F_2(Q^2)$ the residue of the $\Sigma^0$ baryon is needed. The  general
form of the interpolating current for $\Sigma^0$ baryon leads to the following result for its residue  \cite{Rslff15}:

\bea
\label{eslff14}
\lambda_{\Sigma^0}^2 e^{-M_{\Sigma^0}^2/M^2} \es
\frac{1}{256 \pi^4} (5 + 2 \beta + 5 \beta^2 ) M^6 E_2(x) \nnb \\
\ar \frac{m_s}{32 \pi^2} M^2 E_0(x) \left\{ (5 + 2 \beta + 5 \beta^2 ) \qs 
- 6 (-1 + \beta^2 ) \ga \qu+\qd \dr \right\} \nnb \\
\ar \frac{1}{24} \frac{m_0^2}{M^2} (1 - \beta)  \left\{ 6 (1 + \beta) \qs 
\ga \qu + \qd \dr + (-1 + \beta) \qu \qd \right\} \nnb \\
\ar \frac{3 m_s}{32 \pi^2} m_0^2 \ga \qu + \qd \dr (1 - \beta^2 )
\left\{\gamma_E
- \ln \left(\frac{M^2}{\Lambda^2} \right) \right\} \\
\ek \frac{m_s}{192 \pi^2} m_0^2 \left\{ 2 (5 + 2 \beta + 5 \beta^2 ) \qs
- 3 (-1 + \beta^2 ) \ga \qu+\qd \dr \right\} \nnb \\
\ek \frac{1}{6}  (1 - \beta) \left\{ 3 (1 + \beta) \qs \ga \qu + \qd \dr
+ (-1 + \beta) \qu \qd \right\},\nnb
\eea
where
\bea
\label{nolabel}
E_n(x) = 1 - e^x \sum_{k=1}^{n} {x^k\over k!} \nnb
\eea
describes the continuum subtraction and $x=s_0/M^2$. 
It should be noted that the masses and residues of nucleons and other members of the octet baryons, for Ioffe current ($\beta=-1$)
within QCD sum rules approach, were firstly calculated in \cite{ek2,ek3}.

\section{Numerical analysis of the sum rules for the transition form factors}

In order to perform numerical analysis of the transition form factors
$F_1(Q^2)$ and $F_2(Q^2)$ within the light cone QCD sum rules we need to
know the  explicit expressions of the DAs for the $\Lambda$ baryon, as well as
the values of nonperturbative parameters entering into them. These input
parameters for the $\Lambda$ baryon are calculated within the two--point QCD
sum rules method in \cite{Rslff05} which are given as,
\bea
\label{nolabel}
f_\Lambda \es (6.0 \pm 0.3) \times 10^{-3}~GeV^2~, \nnb \\
\lambda_1 \es (1.0 \pm 0.3) \times 10^{-2}~GeV^2~, \nnb \\
\vel \lambda_2 \ver \es (0.83 \pm 0.05) \times 10^{-2}~GeV^2~, \nnb \\
\vel \lambda_3 \ver \es (0.83 \pm 0.05) \times 10^{-2}~GeV^2~. \nnb \\
\eea
Other input parameters used in numerical analysis are $\uu (1~GeV) = \dd
(1~GeV) = - (0.243 \pm 0.01)^3~GeV^3$, $\sp = 0.8\uu$,
$m_0^2(1~GeV)=(0.8\pm0.2)~GeV^2$ \cite{Rslff16}, and $m_{\Sigma_0} = 1.192~GeV$.

Moreover, the sum rules for the transition form factors $F_1(Q^2)$ and
$F_2(Q^2)$ involve the continuum threshold $s_0$, Borel parameter $M^2$
and the arbitrary parameter $\beta$ entering to the expression for the
interpolating current of the $\Sigma_0$ baryon.
For the value of the continuum threshold we shall use
$s_0=(2.8\div 3.0)~GeV^2$, which is obtained from the mass sum rules analysis
\cite{Rslff14}. The Borel parameter $M^2$ is the auxiliary parameter and
physical quantities such as $F_1(Q^2)$ and $F_2(Q^2)$ should be
interdependent of it.
The lower bound of Borel mass is obtained from the condition that the  higher states
and continuum contributions should be less than 40\% of the perturbative
contribution, while the upper limit of $M^2$ is determined by demanding that the light
cone expansion with increasing twist should be convergent. Numerical
analysis shows that both conditions are satisfied when $M^2$ lies in the
region $1.3~GeV^2 \le M^2 \le 2.0~GeV^2$. In our calculations we fix the
lower bound of $Q^2$ to be $Q^2=1.0~GeV^2$, since above this value of $Q^2$
the higher twist contributions are suppressed. In order to guarantee the
higher resonance and continuum contributions to be smaller than the spectral
density contribution, we consider the upper bound of $Q^2$ as $Q^2\le 8.0~GeV^2$. 
In Figs. (1) and (2) we depict the
dependence of the magnetic and electric form factors $G_M(Q^2)$ and
$G_E(Q^2)$ on $Q^2$ at $s_0=3~GeV^2$,  $M^2 = 1.4 ~GeV^2$ and at several fixed values
of $\beta$. From these figures we see that the magnitude of
$G_M(Q^2)$ and $G_E(Q^2)$ for negative (positive) values of $\beta$ are negative
(positive). Only  the $\beta=-1$ case is exceptional and at this value of
$\beta$,
$G_E(Q^2)$ is positive although  its value  is quite small and very
sensitive to the values of the input parameters.

As has already been noted, the sum rules for the transition form factors
$F_1(Q^2)$ and $F_2(Q^2)$ contain also the auxiliary parameter $\beta$. For this
reason we should find the ``working region" of $\beta$, where these form factors
exhibit no dependence on it. For this aim  we shall work with a two-step procedure. At first stage we use the mass sum rules for the $\Sigma^0$ baryon analysis of which leads to 
the domain $-0.6 \le \cos\theta \le0.3$, where  $\beta=\tan\theta$ (see also \cite{Rslff15}). Having this region for $\cos\theta$ obtained from mass sum rules, next, we analyze the 
dependence of form factors on this parameter. Hence,  we present
the dependence of  $G_M(Q^2)$ and $G_E(Q^2)$ on $\cos\theta$ in Figs. (3) and (4) at  several fixed values of
other auxiliary parameters. We see from these figures that  the domain $-0.2 \le \cos\theta \le
0.2$ is the common region where the transition form factors are practically
independent of $\cos\theta$.

In order to compare our predictions on the $Q^2$ dependence of the
transition form factors with the existing ones in the literature we note
that there are only four works \cite{Rslff07,Rslff08,Rslff09,Rslff11} in which
$Q^2$ dependence of the $\gamma^\ast \Lambda \to \Sigma^0$ transition form
factors are studied. In all other works these form factors are studied only
at the point $Q^2=0$. These form factors are studied up to $Q^2=0.4~GeV^2$
in \cite{Rslff11}. Unfortunately the light cone sum rules method is not
applicable in the region $Q^2 \lla 1~GeV^2 \right.$ and for this reason we
can not compare our results with the predictions of \cite{Rslff11}.

When we compare our results on $G_M(Q^2)$ with those given in \cite{Rslff08}
we see that, they are very close to the prediction of \cite{Rslff08}
in the working region of $-0.2\le \cos\theta \le 0.2$, while our results on
$G_E(Q^2)$ are larger compared to those obtained in \cite{Rslff08}.
A comparison of our results on $G_M(Q^2)$ with the ones calculated in
\cite{Rslff09} shows that our predictions are smaller than theirs. However
the situation is contrary  in the case of $G_E(Q^2)$, i.e., our
results are larger compared to the predictions given in \cite{Rslff09}.
Therefore checking the predictions of different approaches on the study of
the $Q^2$ dependence of the form factors for the $\gamma^\ast \Lambda \to
\Sigma^0$ transition receives special interest.
Further improvements of our predictions on the transition form factors
could be achieved by including the ${\cal O}(\alpha_s)$ corrections to DAs, as well as
considering possible future improvements of nonperturbative input
parameters.   

In conclusion, we studied the $\gamma^\ast \Lambda \to \Sigma^0$ transition
form factors within the light cone QCD sum rules using the most general form
of the interpolating current for the $\Sigma^0$ baryon. We obtained the
working regions for the Borel mass parameter and the arbitrary parameter $\beta$
entering to the expressions of the interpolating current. We observed that
the electric charge form factor $G_E(Q^2)$ is quite small as expected. We also
compared our results on $G_E(Q^2)$ and $G_M(Q^2)$ with the predictions existing
in the literature. We saw that our results on $G_M(Q^2)$ are very close to
those that are calculated by the relativistic constituent quark model
\cite{Rslff08}. We further observed that our prediction on the magnetic
(electric charge) form factor is smaller (larger) compared to the results of
the covariant spectator quark model. The $Q^2$ dependence of the transition
form factors presented in this work can be very useful in choosing the right
model.

\newpage

\section*{Figure captions}
{\bf Fig. (1)} The dependence of the magnetic form factor $G_M(Q^2)$ of the
$\gamma^\ast \Lambda \to \Sigma^0$ transition on $Q^2$ at $s_0=3.0~GeV^2$, 
$M^2=1.4~GeV^2$, and at several fixed values of the arbitrary parameter $\beta$. \\ \\
{\bf Fig. (2)} The same as Fig. (1), but for the electric charge form
factor $G_E(Q^2)$. \\ \\
{\bf Fig. (3)} The dependence of the magnetic form factor $G_M$ of the
$\gamma^\ast \Lambda \to \Sigma^0$ transition on $\cos\theta$ at
$Q^2=1.0~GeV^2$,  $s_0=3.0~GeV^2$, and at several fixed values of the
Borel mass parameter
$M^2$. \\ \\
{\bf Fig. (4)} The same as Fig. (2), but for the electric charge form
factor $G_E$.

\newpage

\begin{figure}
\vskip 3. cm
    \includegraphics{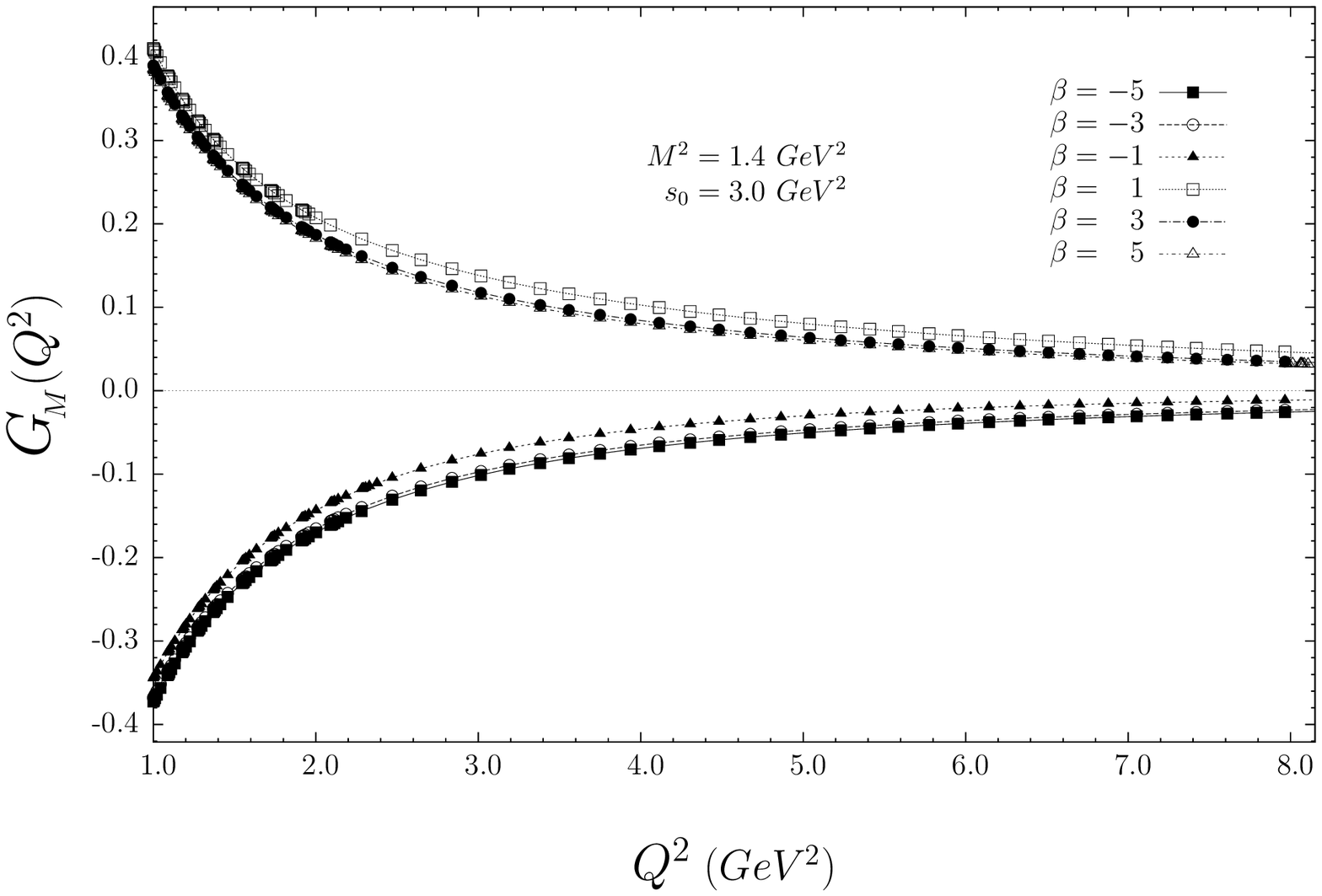}
\vskip 7.0cm
\caption{}
\end{figure}

\begin{figure}
\vskip 4.0 cm
    \includegraphics{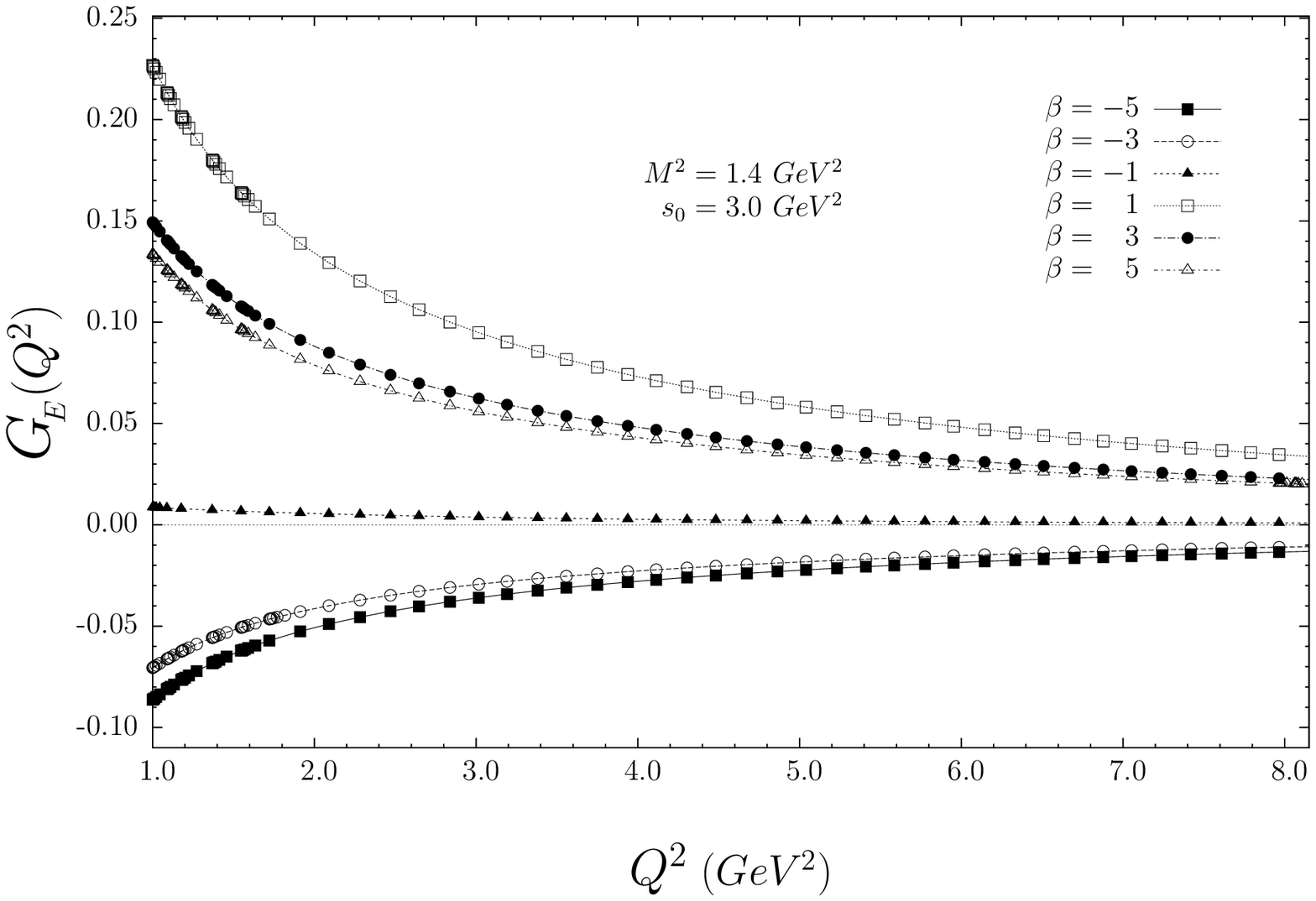}
\vskip 7.0 cm
\caption{}
\end{figure}

\newpage

\begin{figure}
\vskip 3. cm
    \includegraphics{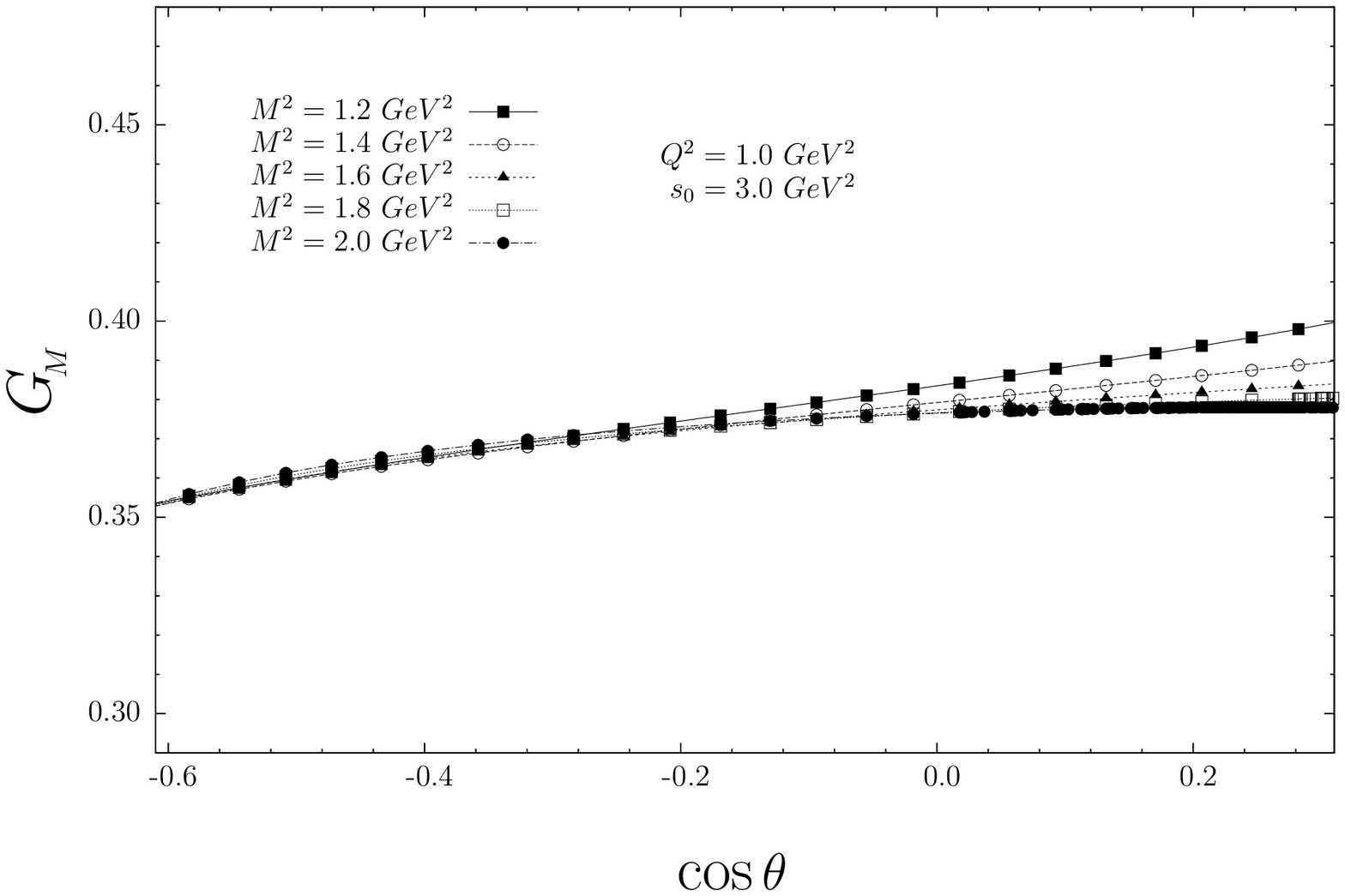}
\vskip 7.0cm
\caption{}
\end{figure}

\begin{figure}
\vskip 4.0 cm
    \includegraphics{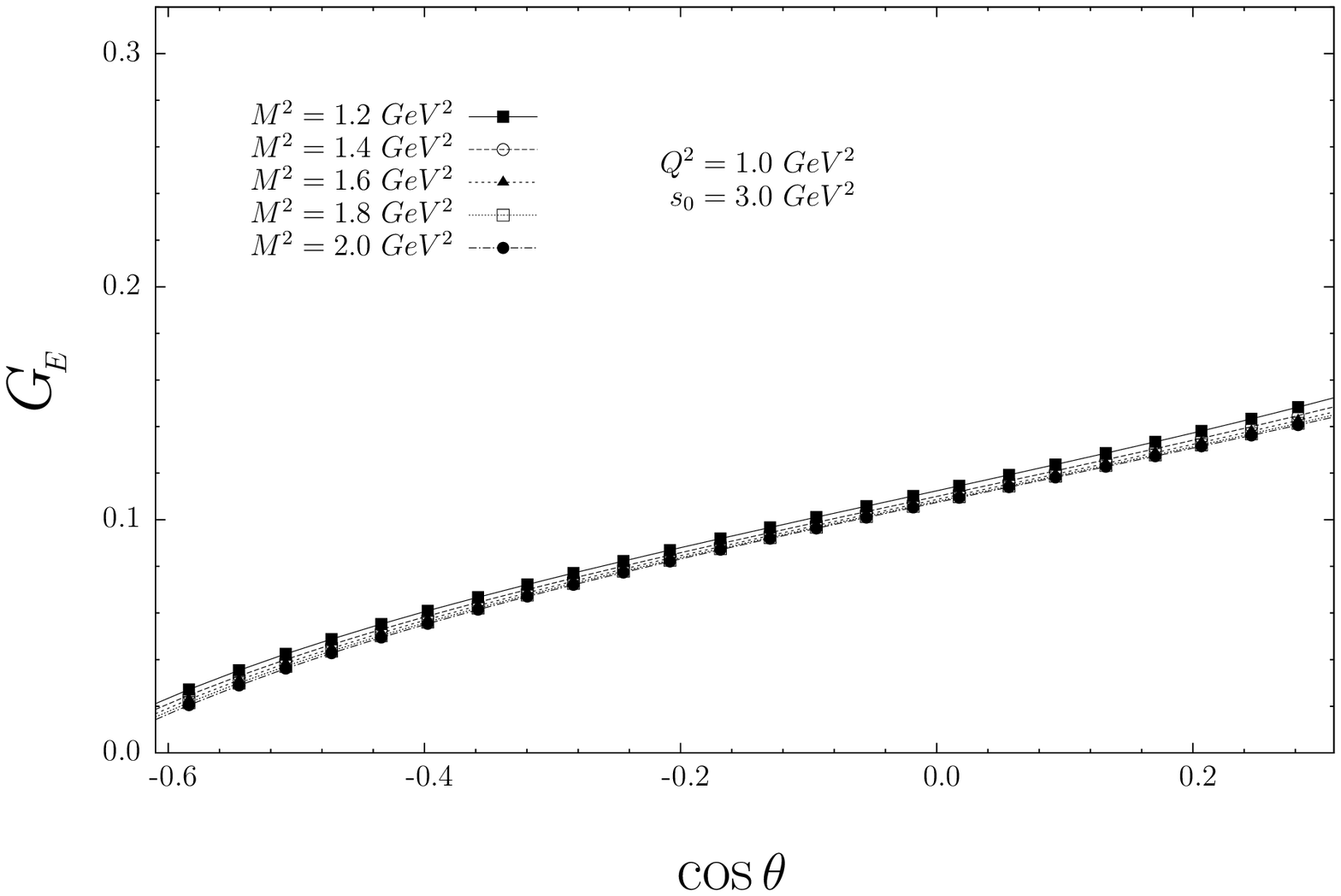}
\vskip 7.0 cm
\caption{}
\end{figure}


\newpage

\begin{thebibliography}{99}

\bibitem{Rslff01} C. F. Perdrisat,
  Prog. Part. Nucl. Phys. {\bf 59}, 694 (2007).

\bibitem{Rslff02} W. K. Brooks, S. Strauch, K. Tsushima,
  J. Phys. Conf. Series {\bf 299}, 012011 (2011).

\bibitem{Rslff03} V. M. Braun, A. Lenz and M. Wittmann,
  Phys. Rev. D {\bf 73}, 094019 (2006).

\bibitem{Rslff04} T. M. Aliev, K. Azizi, A. \"{O}zpineci, and M. Savc{\i},
  Phys. Rev. D {\bf 77}, 114014 (2008).

\bibitem{Rslff05} Yang--Lu Liu, Ming--Qiu Huang,
  Phys. Rev. D {\bf 79}, 114031 (2009);
  Phys. Rev. D {\bf 80}, 055015 (2009).

\bibitem{Rslff06} T. M. Aliev, K. Azizi, M. Savc{\i},
   arXiv:1303.6798 [hep-ph].

\bibitem{Rslff07} H. W. Lin and K. Orginos,
  Phys. Rev. D {\bf 79}, 074507 (2009).

\bibitem{Rslff08} T. Van Cauteren, {\it et. al},
  Eur. Phys. J. A {\bf 20}, 283 (2004).

\bibitem{Morpurgo} G. Morpurgo, Phys. Rev. D {\bf 40}, 2997 (1989); G. Dillon, G. Morpurgo, Phys. Rev. D {\bf 68},  014001 (2003).

\bibitem{Rslff09} G. Ramalho, K. Tsushima,
  Phys. Rev. D {\bf 86}, 114030 (2012).

\bibitem{Rslff10} U. G. Meissner and S. Steininger,
  Nucl. Phys. B {\bf 499}, 349 (1997).

\bibitem{Rslff11} B. Kubis and U. G. Meissner,
  Eur. Phys. J. C {\bf 18}, 747 (2001).

\bibitem{Rslff12} N. Sharma, H. Dahiya, P. K. Chatley and M. Gupta,
  Phys. Rev. D {\bf 81}, 073001 (2010).

\bibitem{Rslff13} N. W. Park and H. Weigel,
  Nucl. Phys. A {\bf 541}, 453 (1992).

\bibitem{Rslff14} T. M. Aliev, A. \"{O}zpineci, M. Savc{\i},
  Phys. Lett. B {\bf 516}, 299 (2001).

\bibitem{Ioffebook} B. L. Ioffe, V. S. Fadin, and L. N. Lipatov, " Quantum Chromodynamics: Perturbative and Nonperturbative Aspects", Cambridge University Press (2012).

\bibitem{Rslff15} T. M. Aliev, A. \"{O}zpineci, M. Savc{\i},
  Phys. Rev. D {\bf 66}, 016002 (2002).

\bibitem{Rslff16} V. M. Belyaev, B. L. Ioffe,
  Sov. Phys. JETP {\bf 56}, 493 (1982).   
\bibitem{ek2} B. L. Ioffe, Nucl. Phys. B {\bf 188}, 317 (1981); Erratum-ibid B {\bf 192}, 591 (1982).
\bibitem{ek3} V. M. Belyaev, B. L. Ioffe, Sov. Phys. JETP {\bf 57}, 484 (1983). 

\end{thebibliography}
\end{document}